# A Broker-based Framework for Integrated SLA-aware SaaS Provisioning

Elarbi Badidi

College of Information Technology, United Arab Emirates University, Al-Ain, United Arab Emirates

## Abstract

*In the service landscape, the issues of service selection, negotiation of Service Level Agreements (SLA), and SLA-compliance monitoring have typically been used in separate and disparate ways, which affect the quality of the services that consumers obtain from their providers. In this work, we propose a broker-based framework to deal with these concerns in an integrated manner for Software as a Service (SaaS) provisioning. The SaaS Broker selects a suitable SaaS provider on behalf of the service consumer by using a utility-driven selection algorithm that ranks the QoS offerings of potential SaaS providers. Then, it negotiates the SLA terms with that provider based on the quality requirements of the service consumer. The monitoring infrastructure observes SLA-compliance during service delivery by using measurements obtained from third-party monitoring services. We also define a utility-based bargaining decision model that allows the service consumer to express her sensitivity for each of the negotiated quality attributes and to evaluate the SaaS provider offer in each round of negotiation. A use-case with few quality attributes and their respective utility functions illustrates the approach.*

## 1. Introduction

Over the last few years, Software as a Service (SaaS) has emerged as one of the most promising service delivery models in cloud computing. The number of SaaS services in various business domains grows continually, and new SaaS service offerings emerge on a steady basis. SaaS is becoming an accepted delivery model for many enterprise applications, including accounting, collaboration, customer relationship management, enterprise resource planning, human resources management, etc. As SaaS services proliferate, users and organizations are becoming more demanding when consuming such services. They increasingly demand services that meet their functional and non-functional requirements. Therefore, SaaS providers need to negotiate Service Level Agreements (SLAs) with their service consumers and adhere to their service level commitments if they want to remain competitive in a changing and demanding business environment.

Given the variety of consumers' requirements, SaaS providers need to manage an increasing number of SLAs, all with potentially changing quality requirements. The SLA lifecycle, as described in the SLA Management Handbook [24], includes six main phases: SLA Establishment, Negotiation and Sales, Implementation, Execution, Assessment, and Decommissioning. Initiatives in standardizing the use of SLAs are (1) WSLA [11] by IBM, (2) the WS-Agreement specification [1] by the Open Grid Forum, and (3) the SLAng Specification Language [14]. Several efforts-- that we discuss in the Related Work section-- investigated the issue of SLA Negotiation [8] [10] [11] [22] [25]. This work shares with these efforts the common goal of providing support for automated SLA negotiation and management. Finding the right SaaS provider is a daunting task for consumers given the abundance and the variety of SaaS offerings. In the service landscape, the critical issues of service-based systems include service discovery, service selection, SLA negotiation, and SLA-compliance monitoring. These concerns were investigated separately using different approaches. We propose in this work an integrated framework to deal with these issues in the context of a SaaS environment. Its principal components are the SaaS Broker and the Monitoring Infrastructure. The SaaS Broker 1) mediates between service consumers and SaaS providers, 2) selects suitable SaaS providers capable of delivering required functionality and quality-of-service, 3) negotiates SLAs on behalf of service consumers, and 4) assesses compliance of service delivery with agreed upon SLA. In each round of negotiation, the SaaS Broker and the selected SaaS provider bargain on multiple SLA parameters by trying to maximize their global utility functions. The Monitoring Infrastructure is in charge of observing service delivery, during SLA implementation, using measurements obtained from independent third party monitoring services.

The remainder of the paper is organized as follows. Next section presents background information on the issues of negotiation and Service Level Agreements in cloud computing. Section 3 highlights related work on SLA negotiation in the context of the Service Oriented Architecture (SOA) and cloud computing environments. Section 4 presents the proposed framework, describes its components and describes our proposed algorithm for QoS-aware SaaS selection and our proposed SLA negotiation model. Section 5 presents a SaaS project management scenario that illustrates the SaaS selection and SLA negotiation. Section 6 concludes the paper and highlights future work.

## 2. BACKGROUND

### 2.1. NEGOTIATION

Negotiation is an important process commonly used in business and personal life to solve conflicts. It aims at reaching a mutual agreement. For instance, negotiations between unions and employers aim at reaching a deal regarding salaries and benefits. It is defined in Wikipedia [30] as: "*Negotiation is a dialogue between two or more people or parties intended to reach a mutually beneficial outcome*".

Negotiation can take different forms and adopt different strategies depending on the multiplicity of the participating parties:

- *One-to-one*: This is the most prevalenttype of negotiation, in which a consumer bargains with a provider for the acquisition of an item or the delivery of a service.
- *One-to-many*: This is a less familiar form of negotiation, in which a consumer bargains with several businesses or providers. These providers can compete or cooperate to share the service delivery to the consumer. An example is when the government negotiates with several telecom operators for mobile service provisioning.
- *Many-to-one*: This is also a less common form of negotiation. In this case, several consumers bargain with a single provider or business. They may compete to reach a deal with the provider or cooperate to share the item or service offered by the provider.
- *Many-to-many*: This form of negotiation involves several consumers and several providers or businesses. A typical example is a negotiation between unions and employers. Companiescooperate to reach an integrated agreement that saves their interests and unions bargain to get the best benefits from employers.

The simplest form of negotiation is a one-to-one negotiation where the two parties bargain on a single issue [15]. However, in practice, the two sides often need to negotiate several issues. For example, a car buyer needs to negotiate with the auto dealer the price, the color, the warranty, the safety options, and the infotainment options. Multi-issue negotiation is notably more complex than single-issue negotiation as a human decision relies on her preference on all issues [13]. A win-win negotiation, in this case, aims to reach a deal that is acceptable to both parties and lets them feel that they have won. An established way to make the negotiation manageable is first to characterize the preferences of each participant with a utility function, and then both parties make decisions based on the evaluation of their respective utility functions. With the proliferation of e-commerce and electronic transactions, automated negotiation is being increasingly used in different multi-agent domains including network bandwidth allocation, robotics, space applications, etc. In these domains, a software agent acts on behalf of a user and negotiates an SLA with a service provider [4] [3].The negotiation process involves three main components:

- *Negotiation objects*: The set of issues the negotiating parties negotiate to reach a mutual agreement. These can include, for example in the case of service provisioning negotiation, the QoS attributes such as availability, response time, reliability, price, and so on.
- *Negotiation protocol*: It determines the cardinality of the participating parties, states their roles, and includes the rules that govern the interaction between the parties. It specifies the negotiation states, the events that cause the change in those states, and the legal actions of the participants in each state. Negotiation protocols fall in general into two classes: bilateral negotiations and auctions. Bilateral negotiations involve two parties, a service provider and a consumer, and protocol for submitting offers and counter-offers until the two sides reach a mutually acceptable agreement on the terms and conditions of business, or one of them withdraws from the negotiation.
- *Decision model*: the decision-making tool that the negotiating parties use to compute their negotiation moves and act in line with the negotiation protocol to achieve their objectives.

## 2.2. SAAS SERVICE LEVEL AGREEMENTS

A Service Level Agreement is anagreement between a service provider and a service consumer consisting of sections concerning the guarantees and commitments to service levels that the service provider will deliver. It describes common understandings and expectations of service between the two parties. The typical components of an SLA are *parties*, *activation-time*, *scope*, *service-level objectives* (SLOs), *penalties*, *exclusions*, and *methods to assess the SLOs* [8]. SLOs represent the goals of the service provider, such as the percentage of service requests the service provider want handled in a specified number of seconds (e.g., 95% within 10 seconds). They represent a commitment of the service provider to maintain a particular level of the service in a predefined period. A typical SLA may have the following SLOs: *service availability*, *system response time*, *service outage resolution time*, and the *reason for an outage*.

Every cloud provider uses its key performance indicators (KPIs). There is no standard at this level yet even though there are some efforts with this regards by the Cloud Services Measurement Initiative Consortium (CSMIC) [21]. Meegan et al. [18] consider that service consumers should expect to have general SaaS indicators in their SLA: cumulative monthly downtime, response time, the persistence of consumer information, and automatic scalability. Burkon L. [2] summarizes the SLA dimensions and their associated quality indicators for the SaaS delivery model. These dimensions are*Availability*, *Response time*, *Throughput*, *Timeliness*, *Reliability*, *Scalability*, *Security*, *Uptime History*, *Granularity*, *Elasticity*, *Interoperability*, *Usability*, and*Testability*.

# 3. RELATED WORK

SLA negotiation and specification of machine-readable SLAs have been the subjects of several efforts in the context of SOA-based environments, computational grid settings, and recently cloud-based environments.

Dan et al. [8] described a framework for providing differentiated service levels to service consumers in an SOA environment using SLAs and automated management. The framework encompasses WSLA – for the creation and negotiation of SLAs --, a system for dynamic allocation of resources based on SLOs, a workload management system that orders requests consistent with the corresponding SLAs, and a system to monitor compliance with the SLA. Chieng et al. [5] described an SLA-driven service provisioning architecture that allows flexible SLA negotiation of services. The emphasis is on bandwidth reservation, which is the most important factor to affect connection's QoS. The negotiation high-level service parameters are price, starting time, session length, and guaranteed bandwidth. Di Modica et al. [9] focused on the use of WS-Agreement for the specification of SLAs and proposed to improve their approach by adding new functionality to the protocol to allow the parties to an agreement to renegotiate and adjust its terms during the service provision.

The ambitious goal of several efforts is to support the automated negotiation of the conditions specified in the SLA. Silaghi et al. [22] introduced a framework for building SLA automatic negotiation strategies under time constraints in computational grids. The framework relies on the recent results from negotiation based on learning opponent strategies that agent-based systems use. They extended the Bayesian learning agent to deal with the limited duration of a bargaining session and showed that opponent learning strategies lead to satisfaction of participants and an optimal allocation of resources. Resinas et al. [20] tackled the problem of building automated service agreement negotiation systems that rely on a bargaining protocol and work in open environments. They proposed a bargaining architecture that provides support for many requirements, which are missing from other proposals, such as multi-term negotiation, heterogeneity of the parties, management of partial information about the parties, and simultaneous negotiations with different parties. Hasselmeyer et al. [10] described a brokered-based approach to SLA negotiation. Their solution relies on the idea of outsourcing SLA negotiation to third parties (agent brokers) acting on behalf of their clients.

An essential component of the European project mOSAIC is the Cloud Agency. The project aims at enabling service consumers to delegate all management tasks, such as SLA management and monitoring of the resources to the agency [28]. The Cloud Agency framework includes several agents that are in charge of managing cloud resources and services provided by various cloud providers.

The SLA@SOI project aims at defining a comprehensive view of SLAs' management and developing a framework for SLA management. A Service Oriented Infrastructure (SOI) can integrate the framework to manage SLA activities [23]. Theilmann et al. [25] developed, in the context of this project, a reference architecture for multi-Level SLA management. This architecture objective is to provide a generic solution for SLA management that can i) cover the complete SLA and service life cycle; ii) support multiple layers SLA management in a service-oriented infrastructure, and iii) be used in various use cases and industrial domains.

The CONTRAIL Project proposed and implemented a cloud federation framework, which aims at relieving the user from managing access to individual service providers [7]. Its federation model, which relies on SLAs, aims at coordinating management and deployment of applications on multiple clouds. Each application could then exploit multiple cloud providers. The user negotiates an SLA with the federation, and the federation negotiates SLAs with one or several providers on behalf of the user. The federation acts as a broker of providers. Selection of providers is SLA-based. Furthermore, SLA specification in Contrail relies on the SLA@SOI syntax.

Macias et al. [16][17] consider a cloud services market in which a client starts negotiating service provisioning with the providers that fulfill its requirements. The client creates an SLA proposal for each provider by filling an SLA template with its requirements. Upon receiving the SLA proposal from the client, each provider evaluates the proposal and returns its SLA offer with pricing information and terms of violations. Afterwards, the client selects the provider whose SLA offer suits best its requirements.

Wu et al. [32] proposed a SaaS broker-based framework for automated SLA negotiation with multiple SaaS providers. SLA bargaining aims to meet customer needs and relies on some heuristics and strategies for generating counter proposals to the SaaS provider's offers. These heuristics take into account some constraints such as market constraints, time and the trade-off between QoS parameters. This work and ours share the same goals and objectives.

Our work shares with these efforts the common goal of providing support for automated SLA negotiation and management. The SaaS Broker with its know-how and value-added services can assist service consumers to (a) express their service requirements (functional and nonfunctional) (b) select appropriate SaaS offerings, (c) negotiate SLA terms, and (d) monitor the implementation of SLAs. However, our approach differs from many of the above approaches, which select a suitable provider after having conducted SLA negotiation with several providers. We consider that the selection of potential SaaS providers should be led first, based on the most recent knowledge on the providers' offerings, and then SLA negotiation has to be carried out only with selected SaaS providers in several rounds of proposals and counter-proposals. We also believe that SLA compliance monitoring should be carried out in collaboration with third-party monitoring services to guarantee the independence of metrics' measurement.

# 4. METHODS

## 4.1. FRAMEWORK OVERVIEW

Figure 1 depicts our proposed integrated framework for SaaS provisioning. The components of the framework are service consumers, the SaaS Broker, the monitoring infrastructure, and SaaS providers.

### 4.1.1. SAAS BROKER

The SaaS Broker is a mediator service that decouples service consumers from SaaS providers. It allows users to subscribe to some services and enables SaaS providers to market their service offerings. Given that service providers and consumers want to focus more on their core businesses rather than negotiating, managing, and monitoring QoS, they delegate management tasks, such as service selection and SLA negotiation, to the SaaS Broker.

The SaaS Broker offers the following management operations: Identity and Access Management, QoS-based SaaSSelection, SLA Negotiation, and Policies Management. Several components implement these operations and cooperate to deliver personalized services to service consumers and SaaS providers. Figure 2 depicts the architecture of the SaaS Broker, which includes the following components: the*Coordinator*, the *Profile Manager*, the *Selection Manager*, the *SLA Manager*, the *Policy Manager*, and the *SLA Monitoring Manager*. The back-end databases maintain information about profiles and preferences of service consumers, SaaS providers' policies, SLAs, and dynamic QoS information. The *Profile Manager* manages the profiles of service consumers, including their preferences for personalized services and QoS. The *Selection Manager* is in charge of implementing policies and strategies for the selection of suitable SaaS providers, based on the service consumer's functional and non-functional requirements and the SaaS providers' QoS offerings. The *SLA Manager* is in charge of carrying out the negotiation process between a service consumer and a SaaS provider to reach agreement as to the service terms and conditions. The *Policy Manager* manages different kinds of policies such as authorization policies, and policies for monitoring services and their QoS. Specific requirements or capabilities of a service provider are declared using XML policy assertion elements. Each assertion describes an atomic aspect of service requirements (e.g. authentication scheme, authorization scheme, QoS characteristics).

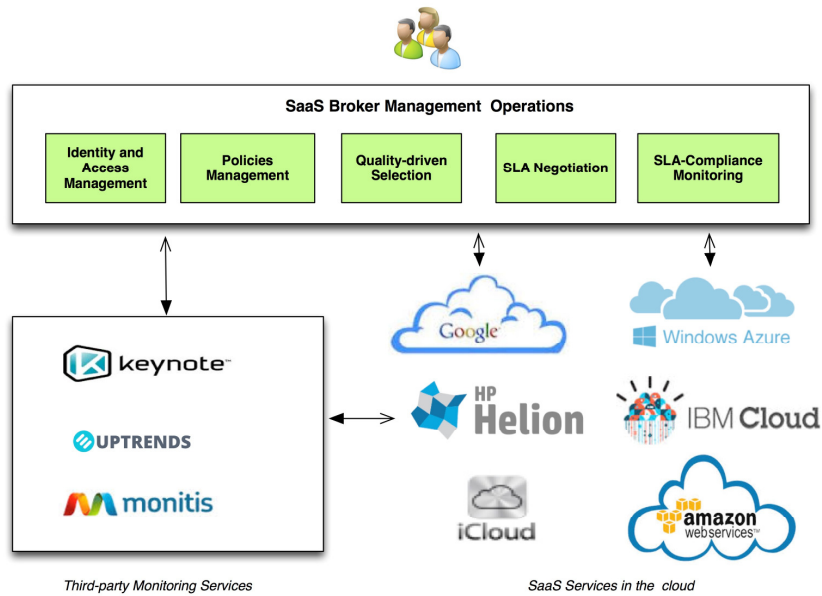

Figure 1. Framework for SLA-driven service provisioning

### 4.1.2. THE MONITORING INFRASTRUCTURE

SaaS services work well most of the time. But, it does not mean that outages and performance issues never happen. Most SaaS providers guarantee availability levels of 99.9% uptime, which translates into about 43 minutes of downtime per month that one can consider as not too much downtime.However, the more a service consumer uses the SaaS service extensively, the more any disruption in service delivery is amplified. The consequences of an interruption in the middle of an important business transaction could be costly.

Many SaaS providers use a "Trust" website, which provides information on service status and disruptions to facilitate accountability for service provision. However, the cause of any disruption in service delivery often comes later after an in-depth assessment of the situation has been made and posted publicly.

Repeated outages or slowdowns might lead service consumers to complain or go away trying to find other alternative providers. With an agreement on a proactive approach to monitoring in real time of the SaaS delivery, the provider can quickly make corrections to mitigate disruptions. Independent performance monitoring from a third-party helps the provider put the required resources to keep the service running at its full potential. It also helps the service consumer understand the level of service and performance obtained from the provider, and, therefore optimize its business processes the SaaS application is supporting. An example of a third-party monitoring service is Keynote [12] with its solution for "Web monitoring for SaaS applications".Other monitoring services includeMonitis[19] and Uptrends [27].

As depicted in Figure 2, the Monitoring Infrastructure includes *Monitoring Plug-ins*that allow interacting with*third-party Monitoring Services*. Each Monitoring Plug-in is in charge of mapping resource metrics, measured by the Monitoring Service, into SLA parameters and monitoring current service levels and their compliance with SLA. Monitoring Plug-ins may be added to the framework if service consumers and SaaS providers agree on using the services of particular monitoring companies. Plug-ins typically use APIs provided by Monitoring Services to get service measurement data. For example, Monitis is providing an open API that allows accessing most of the commands that are available from Monitis dashboard. Similarly, Keynote is offering an API to access Keynote measurement data.

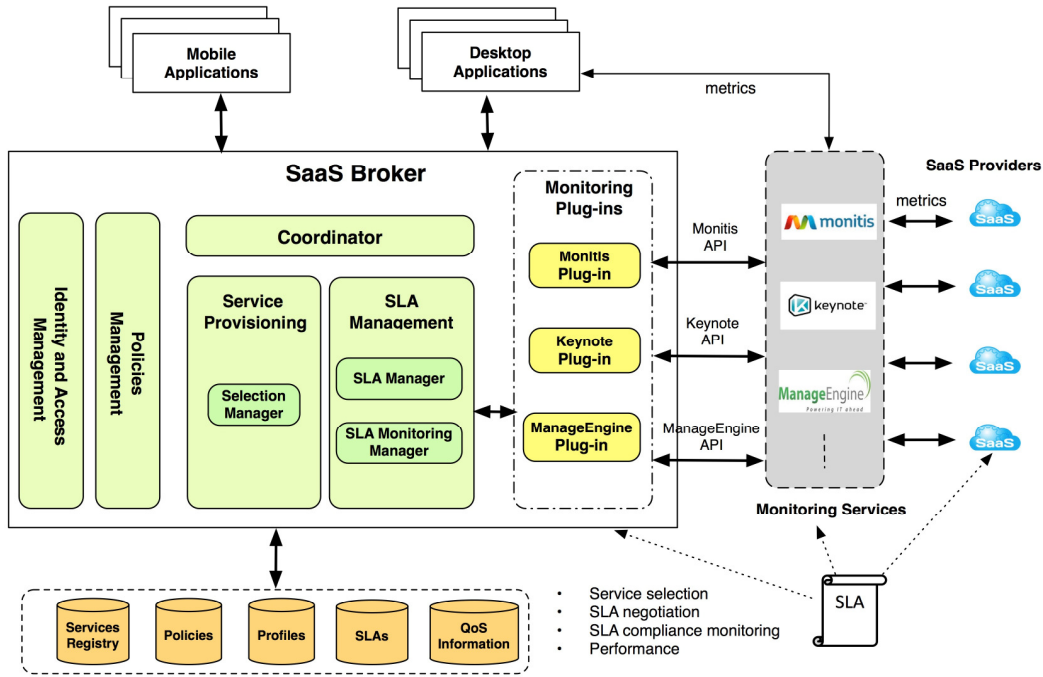

Figure 2. Architecture of the SaaS Broker and the monitoring infrastructure

## 4.2.SLA-BASED SERVICE PROVISIONING

Service provisioning is typically carried out in four phases: Expression of the consumer's requirements and service level expectations, selection of a potential SaaS provider, SLA negotiation and finalization of the contract, service delivery,and SLA compliance monitoring.

### 4.2.1. EXPRESSION OF THE CONSUMER REQUIREMENTS

In this phase, the service consumer submits an SLA Request to the SaaS Broker seeking an appropriate SaaS provider that meets its service functional and non-functional requirements. The request describes the level of service that the service consumer is willing to accept and includes some QoS indicators such as response time, availability, and throughput. After processing the service consumer's authentication, the Coordinator requests its profile from the Profile Manager. If the service consumer's profile is not available in the profile repository, then the Coordinator asks the service consumer to provide additional information, such as service preferences and desired levels of service, to create a new profile for the service consumer. Figure 3 illustrates an example of quality requirements for a project management SaaS service.

```xml
<?xml version="1.0" encoding="UTF-8"?>
...
<service name="ProjectManagementService">
  <QoSAttributes>
    <QoSAttribute>
      <name>Availability</name>
      <min-value>97</min-value>
      <max-value>100</max-value>
      <preferred-value>99.998</preferred-value>
      <unit>percentage</unit>
      <weight>0.3</weight>
    </QoSAttribute>
    <QoSAttribute>
      <name>Response-time</name>
      <min-value>1</min-value>
      <max-value>10</max-value>
      <preferred-value>2</preferred-value>
      <unit>ms</unit>
      <weight>0.2</weight>
    </QoSAttribute>
    <QoSAttribute>
      <name>Reliability</name>
      <min-value>99.95</min-value>
      <max-value>100</max-value>
      <preferred-value>99.996</preferred-value>
      <unit>percentage</unit>
      <weight>0.3</weight>
    </QoSAttribute>
    <QoSAttribute>
      <name>Cost</name>
      <min-value>1</min-value>
      <max-value>40</max-value>
      <preferred-value>20</preferred-value>
      <unit>dollar</unit>
      <weight>0.2</weight>
    </QoSAttribute>
    ...
  </QoSAttributes>
</service>
```

Figure 3. An example of quality requirements specification

### 4.2.2. QoS-aware Service Selection

Before starting the SLA negotiation, the Coordinator requests from the Selection Manager to select appropriate SaaS providers based on the user's QoS requirements and the QoS offerings of SaaS providers advertised with the SaaS Broker or a public registry.

To enter the market, SaaS providers often advertise their service offerings using templates that express the functional and non-functional characteristics of their services. Not all SaaS providers are created equal. They vary by the vertical they are serving, by their maturity, and by the type of application they are offering to their customers. Therefore, service offerings can vary widely regarding service levels and QoS they can deliver. Service descriptions may also vary as they may use different description languages such as WSDL (Web services Description Language) [6], WSOL (Web Services Offering Language) [26], or proprietary languages. The description of their QoS offerings may also be heterogeneous. Various approaches have been used to add QoS support to Web services' description. The first approach advocates incorporating QoS parameters, such as response time, throughput and cost, into the WSDL service description. Other works propose extensions to the Web services Policy Framework to represent QoS policies of Web services. With this heterogeneity in service descriptions, it would be difficult to match services offerings to the user's needs without using a common standard description model (such as WSDL model with QoS support) to map all service offerings to that common model.

A SaaS provider may exceed the service consumer's expectations for some quality attributes while its offering might be below the user's expectations for other quality metrics. Thus, it is imperative to evaluate the overall offering of every potential SaaS provider by taking into account all quality attributes. Such evaluation will allow ranking SaaS providers and selecting the most promising one that can satisfy the consumer request. We describe in the following our proposed algorithm that takes into consideration the QoS requirements of the service consumer and the knowledge of the SaaS Broker on the various service offerings.

Let $Q = \{Q_1, Q_2, \dots, Q_n\}$ be the list of QoS attributes, such as availability, throughput, response time, reputation, and cost of service, required by the service consumer. We classify these QoS attributes into two categories. The first category includes *utility-driven QoS attributes* such as availability and throughput that service consumers would like to maximize. The second category includes *cost-driven QoS attributes* such as response time and cost of the service that service consumers would like to minimize.

Let $SP = \{SP_1, SP_2, \dots, SP_K\}$ be the list of potential SaaS providers, which are capable of providing the type of service requested by the consumer. Let $C = \{C_1, C_2, \dots, C_n\}$, be the vector of quality requirements of the service consumer for each QoS attribute in Q. The following vector expresses the QoS offering of the SaaS provider $SP_r$.

$$Q^r = \{Q_1^r, Q_2^r, \dots, Q_n^r\}$$

The above QoS attributes are expressed in different units and cover a range of values. To be able to sort and rank the SaaS providers' offerings regarding their QoS offerings, we need to normalize values of the QoS offers and define utility functions that map the vector of QoS values into a single real value. We define:

$$Q_i^{max} = \max_{1 \le j \le K}\left(Q_i^j\right)$$
$$Q_i^{min} = \min_{1 \le j \le K}\left(Q_i^j\right)$$

as the maximum and minimum values of the QoS attribute Qi considering all the offers of the potential SaaS providers in SP. The requirement of the service consumer for $Q_i$ may fall in the range $[Q_i^{min}, Q_i^{max}]$ or outside of this range. To take account of the requirements of the service consumer during the normalization, we define the two following values:

$$Q_i'^{max} = max\left(C_i, Q_i^{max}\right)$$
$$Q_i'^{min} = min\left(C_i, Q_i^{min}\right)$$

We define $q_i^r$, the normalized value for the offer of the SaaS provider $SP_r$ regarding the QoS attribute $Q_i$ in equation (1) as:

$$q_i^r = \begin{cases} \frac{Q_i'^{max} - Q_i^r}{Q_i'^{max} - Q_i'^{min}}, & \text{for } cost-driven \text{ QoS attributes} \\ \frac{Q_i^r - Q_i'^{min}}{Q_i'^{max} - Q_i'^{min}}, & \text{for } utility-driven \text{ QoS attributes} \end{cases} \qquad (1)$$

Let $CN = \{c_1, c_2, \dots, c_n\}$, with $0 \le c_i \le 1$, be the vector of normalized values of the quality requirements that the service consumer. When $q_i^r \ge c_i$, it means that the offer of SaaS provider $SP_r$ exceeds the expectation of the service consumer regarding the quality attribute $Q_i$. Otherwise, it means that the offer does not meet the consumer's expectation for $Q_i$. The SaaS Broker may assign (on behalf of the service consumer) an importance weight to each quality attribute. Let $W = \{w_1, w_2, \dots, w_n\}$ be the weight vector associated with the quality attributes

respectively.We define the Utility function (weighted combined level of satisfaction) associated with the offer of SaaS provider$SP_r$as:

$$U_r = \sum_{i=1}^{n} w_i q_i^r \qquad (2)$$

Equation 2 allows ranking the QoS offers of the potential SaaS providers capable of offering the functionality required by the service consumer. The best service offer corresponds to the offer that maximizes the utility function $U_r$.

$$\text{Best service offer} \leftarrow max_{1 \leq r <= K} U_r \qquad (3)$$

Figure 4 summarizes the different steps of the algorithm.Once the Selection Manager has selectedthe most promisingSaaS provider, the SaaS Brokerstarts negotiating SLOs with that SaaS providerthrough its SLA Manager.

**Input**
Q = {Q₁, Q₂, ..., Qₙ} //quality-of-serviceindicators
C = {C₁, C₂, ..., Cⱼ, ..., Cₙ} // quality requirements of the service consumer for each QoS attribute in Q.
W = {w₁, w₂, ..., wₙ, ..., wₙ} //Service consumer weight vector
$SP = \{SP_1, SP_2, ..., SP_K\}$ //list of potential SaaS providers

**Output**
$U = \{U_1, U_2, ..., U_K\}$ // Global utility vector

**Algorithm**
$Q_i^{max} \leftarrow max_{1 \leq j \leq K}(Q_i^j)$
$Q_i^{min} \leftarrow min_{1 \leq j \leq K}(Q_i^j)$
$Q'^{max}_i \leftarrow max(C_i, Q_i^{max})$
$Q'^{min}_i \leftarrow min(C_i, Q_i^{min})$

For each offer $SP_r$ $1 \leq r <= K$
//initialize its global utility to zero
$U_r \leftarrow 0$
//calculate normalized value of each attribute in Q
For each $q_i \in Q$
$q_i^r \leftarrow$ normalized value of qi using Eq. (1)
$c_i \leftarrow$ normalized value of $C_i$ using Eq. (1)

// compute utility function
$U_r \leftarrow U_r + w_i q_i^r$
End For each
End For each

$Best\ offer \leftarrow max(U_r)$ with $1 \leq r <= K$

Figure 1. QoS-aware selection algorithm

### 4.2.3. NEGOTIATION MODEL

The SLA negotiation process of the proposed framework is a bargaining protocol where the SLA Managers of the SaaS Broker and SaaS provider negotiate multiple attributes (QoS attributes). The steps of the SLA negotiation protocol are as follows:

- *Step1*: The SLA Manager of the SaaS Broker forwards the SLA Request to the SLA Manager of the selected SaaS provider requesting an SLA proposal. The SLA request describes only the user's required level of service regarding quality indicators. The SaaS provider parses the SLA Request and validates it against its SLA templates.

- *Step* 2: If the SLA Request is acceptable to the SaaS provider, then its SLA Manager responds to the SLA Request by sending back an SLA proposal to the SaaS Broker. The SaaS Brokeranalyses it to find out if it meets or not to all the service consumer's functional and non-functional requirements. A decision model is used at this point to evaluate the proposal of the SaaS provider.

- *Step* 3: If the SaaS provider can meet the service consumer's expectations, then the SaaS Broker accepts the offer of the SaaS provider and sends an SLA Confirmation to the SLA Manager of the SaaS provider. Otherwise, it rejects the offer and makes a counter-proposal with different conditions, terms, and cost.

- *Step* 4: In the case of the SLA Confirmation, the two parties, service consumer and SaaS provider, approve the agreement, and the service consumer can start using the service according to the terms of the agreement. The agreement specifies the service that the provider should offer to the service consumer, the activation time, the level of QoS to guarantee, the cost of service, the validity period, and the actions to take in the case of a violation of the agreement.

In this process, a multi-issues negotiation takes place between the SLA Managers of both parties, as they have to negotiate multiple QoS attributes concurrently. To reach agreement, The SLA Managers of both parties have to go through several rounds of negotiation of offers and counter-offers until they reach agreement or reach a predefined maximum number of rounds. Figure 5 shows the state charts of the SLA Managers of the negotiating parties. The SaaS Broker, upon reception of an SLA request, maps the service consumer's expectations into normalized QoS values in the range$[0,1]$.

In each round of negotiation, as illustrated in Figure 5, the SLA Manager of the SaaS Broker evaluates the utility function of each QoS attribute and the global utility function to determine whether the offer of the SaaS provider is acceptable or not.Various functions may be used to express the service consumer's utility for eachquality attribute$Q_i$. If $Q_i$ is a utility-driven attribute, such as the availability, its utility function is at its maximal when the SaaS provider can ensure 100% for that quality. If $Q_i$ is a cost-driven attribute, such asresponse time, the corresponding utility function is at its maximal when the SaaS provider can guarantee a lower value, which is very close to zero, for that quality. Let V be the set of utility-driven quality attribute from Q; and Let R be the set of cost-driven quality attribute from Q. $Q = V \cup R$.

We adopt in Equation (4)the following function $F$ to express the utility function of a utility-driven quality attribute X:

$$F(x) = \frac{x^{\beta_X} (1 + \alpha_X)}{1 + \alpha_X x^{\beta_X}} \qquad (4)$$

$x$ is the normalized value of the offer made by the SaaS providerfor X. $\alpha_X$ is the SLA value for X. $\beta_X$represents the sensitivity of the service consumerwith respect to the quality attribute X. When $\beta_X$ is equal to zero, the service consumer is indifferent to the quality attribute X. When $\beta_X$is equal to one, the service consumer is moderately sensitive to the quality attribute X (the relationship is linear). When $\beta_X$is greater than one, the service consumer is increasingly sensitive to the quality attribute X. When$\beta_X$ increases, the service consumer is expressing increasing concern about it. For values of $\beta_X$ smaller than one, the service consumer is expressing increasing indifference for the quality attribute X when $\beta_X$decreases to reachzero.

Similarly, we consider in Equation (5)the function $G$that expresses the utility function of a cost-driven quality attribute Y:

$$G(y) = \frac{1 - y^{\beta_Y}}{1 + \alpha_Y y^{\beta_Y}} \qquad (5)$$

y is the normalized value of the offer made by the SaaS provider for the quality attribute Y. $\alpha_Y$ is the SLA value for Y. $\beta_Y$ represents the sensitivity of the service consumer toward the quality attribute Y. $G$ reaches its maximum, which is 1, when y is 0 and decreases to 0 when y reaches 1.

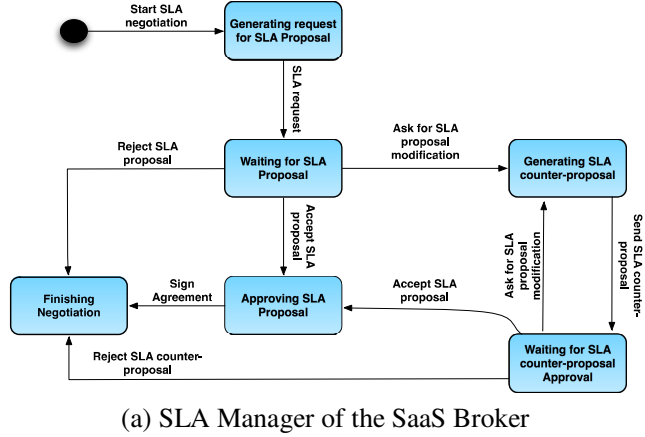

(a) SLA Manager of the SaaS Broker

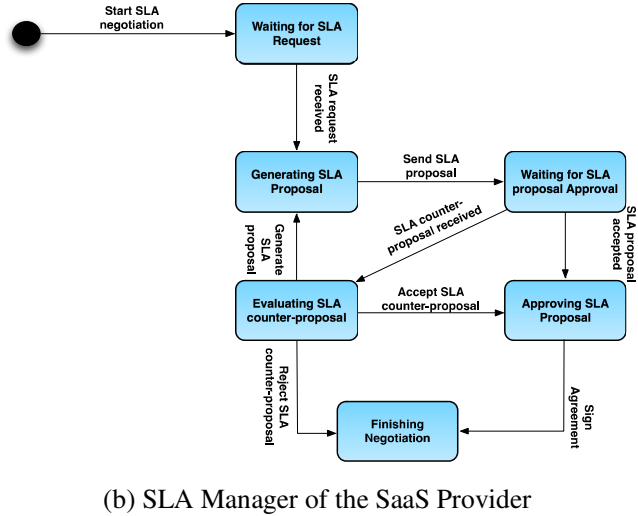

(b) SLA Manager of the SaaS Provider

Figure 5. State chart of the SLA Managers of the negotiating parties.

We define in Equation (6) U as the global utility function, which serves to evaluate the SaaS provider offer during the negotiation phase. We assume that the QoS attributes are independent. Therefore, U can be expressed by the additive linear function of the individual utility functions (F and G) of the quality attributes $Q_i$. $1 \leq i \leq n$

$$U = \sum_{Q_i \in V} w_i F_i + \sum_{Q_j \in R} w_j G_j \qquad (6)$$

$F_i$ represents the utility function associated with the utility-driven QoS attribute $Q_i$. $G_j$ represents the utility function associated with the cost-driven QoS attribute $Q_j$. $w_i$ is the weight that the service consumer assigns to that attribute $Q_i$. Each weight is a number in the range [0,1] and $\sum_1^n w_i = 1$.

# 5. RESULTS

In this section, we consider a scenario from an application domain in which many SaaS offerings are available to businesses. The scenario is used to illustrate the selection process and the negotiation model.

A business company that we call COMPANY would like to enhance its productivity by using project management solutions. On-premise project management tools have limitations when it comes to the mobility of the workforce and the variety of devices that employees may use ranging from desktops to tablets and smartphones. Therefore, a cloud-based solution for project management is increasingly becoming attracting given the way teams may communicate and collaborate, the benefits, and the flexibility it offers to businesses.

Indeed, today businesses have many SaaS solutions for project management support. Although the solutions have many similarities, each solution has its unique features, strengths, and weaknesses. SaaS services for project management are changing the way businesses manage projects. Managers no longer have to print or email on a weekly basis Gantt charts to team members to update them on project status. Besides, the management of multiple projects does not require anymore tracking and exchanging multiple Microsoft Project documents. Examples of SaaS solutions for project management are *basecamp*, *GroupCamp*, *Planzone*, *Clarizen*, *Zoho project*, and *Beesy*. Table 1 presents the main features of these SaaS offerings.

Table 1- SaaS Project Management Solutions

| Operation \SaaS PM solution | basecamp | GroupCamp | Planzone | Clarizen | Zoho project | Beesy |
|---|---|---|---|---|---|---|
| Defining tasks | ✓ | ✓ | ✓ | ✓ | ✓ | ✓ |
| Assigning resources | ✓ | ✓ | ✓ | ✓ | ✓ | ✓ |
| Specifying milestones | ✓ | ✓ | ✓ | ✓ | ✓ | ✓ |
| Generating detailed reports | ✓ | ✓ | ✓ | ✓ | ✓ | ✓ |
| Parallel tasks support | | ✓ | | | | |
| Sending email to create new task | ✓ | ✓ | | ✓ | ✓ | ✓ |
| Templates to specify project data structures (team members, tasks, wiki page, etc..) | | | ✓ | ✓ | ✓ | |
| Advanced reporting | | ✓ | | ✓ | ✓ | |
| Pricing scheme | Monthly subscription & yearly subscription | Six pricing plans: (Free, Start, Plus, Best, Max) | Four monthly plans (Basic, Team, Business, Enterprise) & customized plan. | Three Monthly plans ( Professional/ Enterprise/ Unlimited) | Four monthly plans (Free, Express, Premium, Enterprise) | Four monthly plans (Free, Sync, Pro, Team) |
| SLA support | Yes for yearly subscription | | | | SLA management | |

Selecting a SaaS solution by COMPANY is not an easy task given the plethora of features, pricing schemes, and SLAs. COMPANY has first to understand and specify its project management needs. Then, it can delegate the selection of an appropriate SaaS solution to the SaaS Broker, who has better knowledge on the features of each SaaS solution. The selection process starts by considering the solutions that might satisfy the functional requirements of COMPANY. Then, potential SaaS offerings will be ranked based on its non-functional requirements. In this first phase, high-level requirements, such as the scope of COMPANY's project management needs, might be used to identify candidate services. The scope may be project specific, department level, or enterprise level. The selection process might then be refined with other detailed or low-level functional requirements. Assuming that the SaaS Broker has knowledge on the non-functional offerings of the potential SaaS providers, our proposed algorithm might be used to rank them.

We assume that the QoS requirement of the service consumer are: *Availability=99.97%, Reliability=99.96%, Cost=25$, and Response time=6ms*. Table 2 shows the QoS values offered by 24 potential SaaS providers. Table 3 lists the weights that the SaaS Broker assigns to these QoS parameters according to the consumer requirements. By using our selection algorithm, normalized values are calculated for each QoS parameter while differentiating between utility-driven QoS attributes and cost-driven QoS attributes. Then, using the weight table, the aggregate utility function is calculated for each SaaS offering. Figure 6 depicts the normalized values of each of these QoS attributes.

Table 2- Qos Offerings of SaaS Providers

| SaaS_ID | Availability | Reliability | Cost ($) | Response time (ms) |
|---------|--------------|-------------|----------|--------------------|
| 1 | 0.99988 | 0.9995 | 16.1 | 6 |
| 2 | 0.99968 | 0.99953 | 38.1 | 2 |
| 3 | 0.99935 | 0.99962 | 8.4 | 3 |
| 4 | 0.99988 | 0.99964 | 40.2 | 3 |
| 5 | 0.99959 | 0.99954 | 12.6 | 4 |
| 6 | 0.99963 | 0.99958 | 22.2 | 6 |
| 7 | 0.99939 | 0.99971 | 33.2 | 7 |
| 8 | 0.99918 | 0.99975 | 25.3 | 2 |
| 9 | 0.99995 | 0.9999 | 30.8 | 7 |
| 10 | 0.99958 | 0.99956 | 24.2 | 3 |
| 11 | 0.99945 | 0.99971 | 22.8 | 7 |
| 12 | 0.99981 | 0.99976 | 6.7 | 7 |
| 13 | 0.99911 | 0.99987 | 15 | 3 |
| 14 | 0.99924 | 0.99983 | 11.7 | 5 |
| 15 | 0.99912 | 0.9998 | 24.8 | 5 |
| 16 | 0.99948 | 0.99973 | 22.7 | 3 |
| 17 | 0.99952 | 0.99967 | 31.9 | 5 |
| 18 | 0.99999 | 0.99962 | 23.9 | 2 |
| 19 | 0.99944 | 0.99975 | 33.7 | 3 |
| 20 | 0.99943 | 0.99972 | 20.7 | 5 |
| 21 | 0.99987 | 0.99957 | 19.6 | 5 |
| 22 | 0.99959 | 0.99977 | 27.2 | 2 |
| 23 | 0.9997 | 0.99952 | 20.4 | 4 |
| 24 | 0.99999 | 0.99992 | 25.4 | 5 |

Table 4 depicts the rank table generated by the SaaS Broker using our proposed algorithm. The offer of the SaaS provider with ID 24 is the best offer as it maximizes the aggregated utility. In parallel with that, we also use the TOPSIS method [31] for ranking the SaaS offerings. We identified first the ideal solution (IS) and the negative ideal solution (NIS). Then, we calculated for each SaaS offer the separation from IS from the NIS. Afterwards, we calculate the closeness to the ideal solution (CIS). Table 5 depicts the ranking using the TOPSIS method.

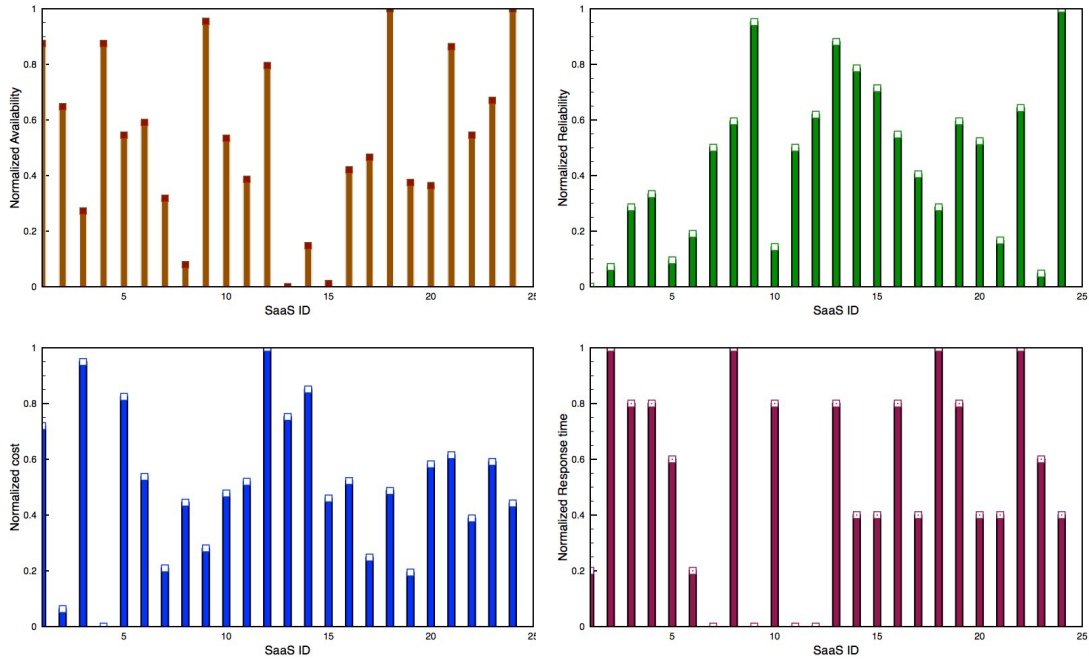

Figure 6. Normalized values of the QoS attributes

Table 3- Weighted table of the QoS attributes

|  | Availability (%) | Reliability (%) | Cost ($) | Response time (ms) |
|---|---|---|---|---|
| **Weight** | 0.305 | 0.267 | 0.197 | 0.231 |

The results show that both algorithms have obtained the same best SaaS provider (SaaS ID: 24). Subsequent SaaS offerings in the two rankings are very close, especially for the first ten best offerings. This process results in identifying a suitable SaaS provider for project management. Reaching an agreement with that provider would be the next phase before COMPANY would be able to use the SaaS solution for its project management activities.

To illustrate our negotiation model, we consider that negotiation would be mainly about two quality attributes: availability and response time. In this scenario, both availability and response time are important qualities for COMPANY. The desired level of the availability is 99.97% uptime and 20% (6ms) for the response time ($\alpha_V = 0.99$ and $\alpha_R = 0.20$). In addition, COMPANY gives 70% importance weight to the availability and 30% weight to response time ($w_V = 0.70$ and $w_R = 0.30$).

Figure 7 shows the evolution of the utility function associated with Availability for different values of $\beta_V$, the sensitivity of COMPANY to the availability of the service ($\beta_V = 1$, 2, and 4 respectively). At low offers of availability by the SaaS provider, the utility is also low. As the SaaS offer increases, the utility rapidly increases when COMPANY is not sensitive to the availability ($\beta_V = 1$), moderately when $\beta_V = 2$, and slowly as availability becomes a sensitive issue for the business ($\beta_V = 4$).

Table 4- Ranking of SaaS Offerings using our proposed algorithm

| SaaS_ID | Utility (%) | Rank |
|---|---|---|
| 24 | 65.90 | 1 |
| 12 | 60.49 | 2 |
| 9 | 60.07 | 3 |
| 18 | 47.71 | 4 |
| 21 | 42.90 | 5 |
| 14 | 42.24 | 6 |
| 22 | 41.45 | 7 |
| 1 | 40.86 | 8 |
| 13 | 38.34 | 9 |
| 16 | 37.74 | 10 |
| 20 | 36.54 | 11 |
| 4 | 35.59 | 12 |
| 5 | 35.41 | 13 |
| 11 | 35.37 | 14 |
| 3 | 34.65 | 15 |
| 6 | 33.69 | 16 |
| 23 | 33.36 | 17 |
| 19 | 31.15 | 18 |
| 17 | 29.90 | 19 |
| 10 | 29.51 | 20 |
| 15 | 28.47 | 21 |
| 7 | 27.17 | 22 |
| 8 | 27.08 | 23 |
| 2 | 22.90 | 24 |

Table 5- Ranking of SaaS Offerings using the TOPSIS method [31]

| SaaS_ID | CIS (%) | Rank |
|---|---|---|
| 24 | 70.60 | 1 |
| 18 | 65.05 | 2 |
| 22 | 62.23 | 3 |
| 9 | 58.97 | 4 |
| 12 | 57.62 | 5 |
| 4 | 55.29 | 6 |
| 16 | 54.66 | 7 |
| 21 | 52.70 | 8 |
| 13 | 51.54 | 9 |
| 19 | 49.76 | 10 |
| 3 | 49.12 | 11 |
| 14 | 48.81 | 12 |
| 2 | 48.22 | 13 |
| 1 | 47.97 | 14 |
| 8 | 47.94 | 15 |
| 5 | 47.91 | 16 |
| 23 | 47.71 | 17 |
| 10 | 47.71 | 18 |
| 20 | 45.03 | 19 |
| 6 | 40.47 | 20 |
| 17 | 40.42 | 21 |
| 15 | 39.25 | 22 |
| 11 | 37.70 | 23 |
| 7 | 31.35 | 24 |

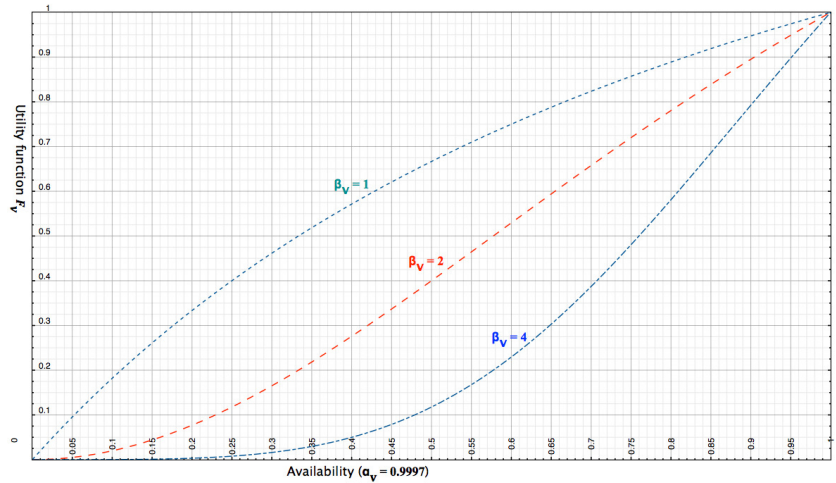

Figure 7. Utility function of the availability ($F_V$) with different values of the sensitivity factor $\beta_V$

Similarly, Figure 8 shows the evolution of the utility function associated with response time for different values of $\beta_R$, the sensitivity of COMPANY to the response time quality attribute ($\beta_R = 1$, 2, and 4 respectively). At low response time offers by the SaaS provider, the utility is high. As the SaaS offer for response time increases, the utility rapidly decreases when COMPANY is not sensitive to the response time ($\beta_R = 1$), moderately when $\beta_R = 2$, and slowly when response time is a sensitive issue for the business ($\beta_R = 4$).

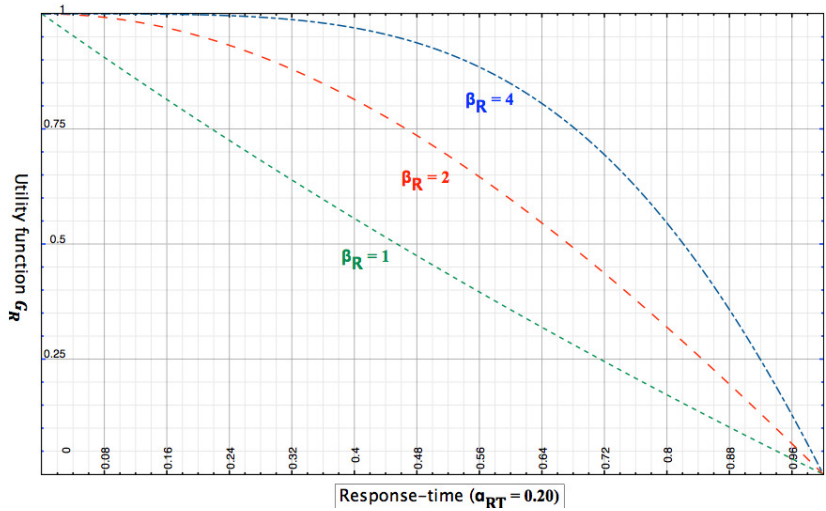

Figure 8. Utility of the response time ($G_R$) with different values of the sensitivity factor $\beta_R$

Assuming that COMPANY is more sensitive to availability than to response time (**$\beta_V$**= 4 and **$\beta_R$**= 2).Figure 9 plots the additive utility function for this scenario. It shows that the overall utility function reaches its maximal value when the response time is low (close to 0), and the availability is high (close to 1).The SaaS Broker, acting on behalf of COMPANY, may set a threshold for the overall utility function. When the offer of the SaaS provider is below that threshold, the SaaS Broker rejects the offer and sends a counter-proposal to the SaaS provider.

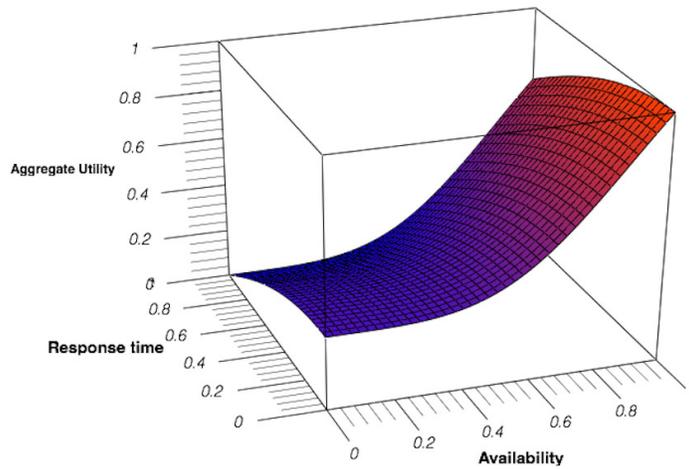

Figure 9.Overall utility function of the service consumer

## 6. CONCLUSION

With the rising adoption of Software-as-a-Service as a service provisioning model, SaaS providers need to implement and manage a growing number of SLAs to remain competitive in a highly demanding business environment. On the other hand, service consumers need to find appropriate SaaS providers that meet their functional and non-functional requirements and negotiate with them the terms of service delivery.

In this paper, we have presented a framework for SLA-based service provisioning, which relies on a SaaS Broker for a) mediating between service consumers and SaaS providers, b) selecting suitable SaaS providers that can meet service consumers' expectations, and c) negotiating the terms of the SLA on behalf of service consumers. SLA negotiation involves the negotiation of multiple QoS attributes with both parties trying to maximize their utility function in several rounds of proposals and counter-proposals. We proposed a SaaS selection algorithm and a decision negotiation model that the SaaS Broker may implement to evaluate the offers of the selected SaaS provider in each round of negotiation. We have considered a scenario with two quality attributes, availability and response time, to illustrate the approach. The proposed decision model allows the consumer to specify her sensitivity for each negotiated quality attribute. Our approach differs from other approaches that carry out SLA negotiation with several cloud service providers, and then select the one with the best offer. It first selects the best suitable SaaS provider based on prior knowledge of the SaaS Broker on the providers' offerings. Then, it negotiates with the selected SaaS provider the SLA terms in several rounds of proposals and counter-proposals until an agreement or a timeout is reached.

As future work, we intend to investigate how the SLA negotiation evolves when the importance weights, associated with the QoS attributes, vary in time and, then, study the effect of the interdependence between QoS attributes on the SLA negotiation. Also, we intend to build a prototype of our proposed framework.